# Computing with injection-locked spintronic diodes

Luciano Mazza[1], Vito Puliafito[1,*], Eleonora Raimondo[2], Anna Giordano[3], Zhongming Zeng[4], Mario Carpentieri[1], Giovanni Finocchio[2,*]

[1] Department of Electrical and Information Engineering, Politecnico di Bari, via Orabona 4, 70125 Bari, Italy

[2] Department of Mathematical and Computer Sciences, Physical Sciences and Earth Sciences, University of Messina, I-98166, Messina, Italy

[3] Department of Engineering, University of Messina, c.da Di Dio, 98166 Messina, Italy

[4] Suzhou Institute of Nano-tech and Nano-bionics, Chinese Academy of Sciences, Ruoshui Road 398, Suzhou 215123, P. R. China 398, Suzhou 215123, P. R. China

Corresponding author: *vito.puliafito@poliba.it, *gfinocchio@unime.it

**Abstract**

Spintronic diodes (STDs) are emerging as a technology for the realization of high-performance microwave detectors. The key advantages of such devices are their high sensitivity, capability to work at low input power, and compactness. In this work, we show a possible use of STDs for neuromorphic computing expanding the realm of their functionalities to implement analog multiplication, which is a key operation in convolutional neural networks (CNN). In particular, we introduce the concept of degree of rectification (DOR) in injection-locked STDs. Micromagnetic simulations are used to design and identify the working range of the STDs for the implementation of the DOR. Previous experimental data confirm the applicability of the proposed solution, which is tested in image processing and in a CNN that recognizes handwritten digits.

# I. INTRODUCTION

The use of deep convolutional neural networks (CNNs) has grown exponentially impacting the development of the emerging area of artificial intelligence. The realization of low power and compact building blocks (neurons and synapses) for such networks is a key requirement for their hardware realization and massive integration in consumer electronics and Internet of Things nodes [1].

One of the most critical operations for CNNs is multiplication, which is implemented to calculate the inputs of each neuron and to perform the convolution operation needed for features extraction.

Digital multiplication is realized with multiple adders, multiplexers and carry-over systems. This approach is scalable but critical in terms of energy dissipation, computing time, and area occupancy [2].

An analog electronic implementation of the convolution may overcome today's bottlenecks for the realization of an efficient device in terms of computational time and energy consumption, but it turns out to be very difficult to obtain because of the susceptibility to noise and voltage offsets [3] [4].

A possible direction for analog solutions is the use of photonic tensor cores based on phase-change materials [5]. This approach is parallelizable since the inputs may be multiplexed in the frequency domain, but because of its size it is not suitable for compact systems. Another solution is based on memristors [6] [7] where the multiplication is obtained using Ohm's law. For applications in CNNs, however, the main problems of memristors are the device-to-device variability and conductance degradation, which could be corrected with continuous in-situ training of the CNN weights [7].

From a theoretical point of view, the analog multiplication between two values $F$ and $G$ can be also implemented considering an observable $P$ characterized by a parabolic input-output relation $P(X) = \mathrm{a}X^2 + \mathrm{b}X + \mathrm{c}$, where a, b, and c are characteristic parameters of the physical system and $X$ is the input. Combining three measurements, where $X = F - G$, $X = F$, and $X = -G$, it can be easily demonstrated that the multiplication $FG$ is given by:

$$FG = \frac{P(F-G) - P(F) - P(-G) + c}{-2a} \tag{1}$$

A description of the use of Eq. 1 is presented in Section IV.

The degree of match (DOM) between two interacting nonlinear oscillators is defined as:

$$DOM(t) = \frac{1}{2}|z_1(t) + z_2(t)| \tag{2}$$

where $z_1(t)$ and $z_2(t)$ are the complex oscillating variable describing the behavior of each oscillator which is characterized by a power $p = |z|^2$ and phase $\phi = arg(z)$. As already demonstrated theoretically in previous works, couples of tunable CMOS (Complementary Metal-Oxide

Semiconductor) [8] and spintronic [9] oscillators exhibit a time-independent DOM in the locking region which can be used for the computation of multiplication by using Eq. 1.

In particular, the calculation of the DOM using spintronic devices seems to be promising in terms of time and energy [10], compactness (nanoscale size [11]) and CMOS compatibility (spin-transfer-torque MRAM (Magnetoresistive Random Access Memory) have already been integrated with the CMOS processes by the main foundries [12]). For two equal spintronic oscillators having an output power approximately constant in the locking region, the DOM is given by:

$$DOM \cong \left|\sqrt{p}\right| \cos \frac{\Delta\phi}{2} \qquad (3)$$

where $\Delta\phi = \phi_1 - \phi_2$ is the phase difference between the time domain traces of the two oscillators [8] [9]. When in the working point of the oscillators, which can be set and controlled by a bias current [9], $\Delta\phi$ is close to 0, the cosine can be approximated by a second order polynomial function applying the Taylor expansion, therefore the DOM can be eligible as observable $P$ in Eq. 1 ($X$ being the oscillator phase difference). Similar remarks can be generalized in the case of two different oscillators [13]. However, the read-out mechanism of the DOM is a key problem because the amplitude and phase of each oscillator are not easy for on-chip evaluation.

Recent developments of spintronic diodes (STDs) gave rise to room temperature solutions overcoming the thermodynamic limits of Schottky diodes [14] [15] [16] [17] [18].

The self-oscillation state of the magnetoresistive signal in STDs is driven by a sufficiently large dc current. When the device is locked to an external ac current, it gives rise to a rectification voltage $V_{dc}$ with sensitivities exceeding 0.2 MV/W [19] [20], which concretely means that an order of mV of rectification voltage can be achieved at an input power of tens of nW or less. This sensitivity has been recently improved by an order of magnitude by combining a bolometric effect with the injection locking [20].

This work introduces the concept of degree of rectification (DOR) which is based on the idea to use Eq. 1 to compute the analog multiplication having the rectified voltage $V_{dc}$ of injection-locked STDs as an observable $P$ and the bias current $I_{dc}$ as input. The applicability of DOR is studied with micromagnetic simulations, which show the fundamental aspects originating the parabolic relationship between $V_{dc}$ and $I_{dc}$ [19], and with experimental data from ref. [15]. Such an approach has the intrinsic advantages of spintronic technology (CMOS compatible, scalable, and low power dissipation), simplifies the readout mechanism as compared to the DOM, and can be realized with a single device. To test the robustness of the DOR based multiplication, we have realized in software a simple CNN that recognizes the digits of the MNIST (Modified National Institute of Standards and Technology) dataset [21] [22] where the convolution is obtained with DOR based and ideal multiplication. Then we have emulated the effect of transfer learning in hardware by computing the performance of the network, in terms of recognition accuracy, for the same dataset. The results show that the transferring of knowledge is robust and, introducing the DOR multiplication in the training phase of the fully connected (FC) layer of the CNN, the accuracy of the ideal CNN is recovered. We show that the use of DOR is a reliable method for the extraction of the dark knowledge used for knowledge distillation [23] [24], and it is also robust to thermal fluctuations and device-to-device variability.

## II. DEVICE CONCEPT AND MICROMAGNETIC MODELING

The device idea is a hybrid Magnetic Tunnel Junction (MTJ), sketched in Fig. 1(a), having an out-of-plane free layer (FL, 1.63 nm-thick $Co_{20}Fe_{60}B_{20}$) and an in-plane polarizer (PL, synthetic antiferromagnet $Co_{70}Fe_{30}$ (2.3 nm)/Ru (0.85 nm)/$Co_{40}Fe_{40}B_{20}$ (2.4 nm)) exchange biased by a PtMn (15 nm) layer. The device is patterned with an elliptical cross-section (150 nm × 60 nm) and its resistances in the parallel and antiparallel states are $R_P = 640\ \Omega$ and $R_{AP} = 1200\ \Omega$, respectively. The further advantage of this device is the zero-field operation as already demonstrated [25].

To explain the concept of DOR, we performed micromagnetic simulations of the MTJ's FL magnetization solving numerically the Landau-Lifshitz-Gilbert-Slonczewski (LLGS) equation [15] [25] [26]:

$$(1 + \alpha^2)\frac{d\mathbf{m}}{d\tau} = -(\mathbf{m} \times \mathbf{h}_{\text{eff}}) - \alpha \mathbf{m} \times (\mathbf{m} \times \mathbf{h}_{\text{eff}}) + \\ + \sigma I g_T[\mathbf{m} \times (\mathbf{m} \times \mathbf{m}_\text{p}) - q(\mathbf{m} \times \mathbf{m}_\text{p})] \quad (4)$$

On the right side of Eq. 4, the three terms represent the conservative dynamics, the Gilbert dissipation, and the spin-transfer torque. $\alpha$ is the damping parameter, $\mathbf{m} = \mathbf{M}/M_S$ is the normalized magnetization of the FL, $M_S$ being its saturation magnetization; $d\tau = \gamma_0 M_S dt$ is the dimensionless time step, where $\gamma_0$ is the gyromagnetic ratio. The effective field $\mathbf{h}_{\text{eff}}$ includes the standard micromagnetic contributions from the exchange, anisotropy, external, and demagnetizing field. The torque term is proportional to $\sigma = \frac{g|\mu_B|}{|e|\gamma_0 M_S^2 V_{FL}}$, where $g$ is the gyromagnetic splitting factor, $\mu_B$ is the Bohr's magneton, $e$ is the electron charge, $V_{FL}$ is the FL volume. $I$ is the total current flowing into the MTJ given by $I = I_{dc} + I_{ac,MAX} \sin(2\pi f_{ac} + \varphi_{ac})$. For the scalar function $g_T$, we have considered the expression $g_T(\theta) = 2\eta_T[1 + \eta_T^2 \cos(\theta)]^{-1}$, where $\eta_T$ is the polarization efficiency and $\cos(\theta) = \mathbf{m} \cdot \mathbf{m}_p$ [26] [27] with $\mathbf{m}_p = \mathbf{M}/M_{SP}$ the normalized magnetization of the PL and $M_{SP}$ its saturation magnetization. The micromagnetic parameters are $M_S = 9.5 \times 10^5$ A/m, perpendicular anisotropy constant $k_U = 5.45 \times 10^5$ J/m$^3$, exchange constant $A = 2.0 \times 10^{-11}$ J/m, $\alpha = 0.02$, and $\eta_T = 0.66$. The spin-transfer-torque includes both damping-like and field-like torques, where $q$ is the ratio between the two torque contributions and it is equal to 0.1 [28], the results achieved for $q = 0$ and 0.2 are qualitatively the same with a variation in the oscillation frequency of less than 3%. $\mathbf{m}_p$ is aligned along the -x direction.

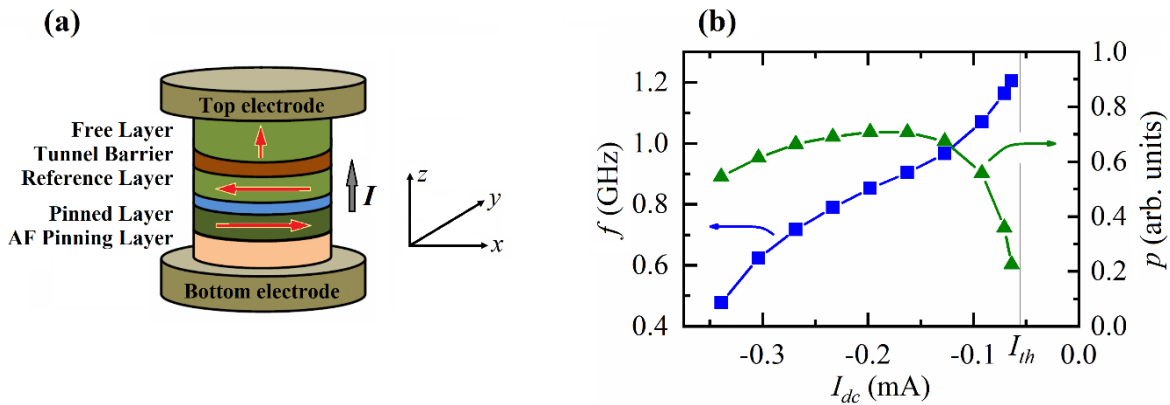

FIG. 1. (a) A sketch of the STD with the indication of the free and polarizer layer. The structure is hybrid, the free layer is perpendicular, and the polarizer is aligned along the -x direction. The Cartesian coordinate system is also included as an inset. (b) Frequency (blue squares) and power (green triangles) of the FL magnetization oscillation as a function of the applied dc current, obtained by means of micromagnetic simulations. The vertical line represents the threshold current $|I_{th}| = 0.056$ mA.

## III. SIMULATION RESULTS

First, we have characterized the properties of self-oscillations, frequency $f_0$ (blue squares) and power $p_0$ (green triangles), as a function of the dc current which are summarized in Fig. 1(b). We found a threshold current $|I_{th}| = 0.056$ mA and a nonlinear frequency shift $N/2\pi = df_0/dp_0 \cong -411$ MHz; the latter is a parameter that links the oscillation frequency and power of a free-running spin-transfer-torque oscillator near the threshold current $f_0 = f_0(I_{th}) + Np_0/2\pi$ [29]. As discussed ahead in the text, $N$ is a crucial parameter for the properties of the magnetization dynamics in the injection locking regime which is achieved when $f_{ac}$ approaches the self-oscillation frequency $f_0$. In the locking regime, the magnetoresistance oscillates at the same frequency of the ac current hence it is possible to observe a rectification voltage [30] [31]. The variation of the input dc current, within a specific locking range, does not lead to a change of the frequency, but it can modify the amplitude of the oscillating magnetization $dm_X(I_{dc})$, which is linked to the oscillator power $p$ ($dm_X = \sqrt{p}$), and the intrinsic phase shift $\varphi(I_{dc})$ between the ac current and the oscillating magnetoresistive signal [32]. The output voltage can be computed with the following expression [15]:

$$V_{dc} = \frac{(R_{AP}-R_P)\sqrt{p}}{4} I_{ac,MAX} \cos(\varphi(I_{dc})) \qquad (5)$$

Fig. 2(a) shows an example of rectification voltage obtained for $I_{ac,MAX} = 70.7$ µA, $f_{ac} = 800$ MHz and $\varphi_{ac} = 0$. The rectification voltage has a maximum at about $I_{dc,0} = -0.134$ mA corresponding to a phase shift close to 0 (see Fig. 2(b), where the additional phase shift introduced by the direction of the polarizer has not been taken into account). Fig. 2(b) shows $dm_X$ (blue diamonds) and $\varphi$ (green squares) for the simulations reported in Fig. 2(a). The intrinsic phase shift is a quasi-linear function of the dc current with a slight deviation close to the edge of the locking region similar to what was also observed in [32] [33]. Instead, the amplitude of the magnetization is weakly dependent on the current. This result is expected for an oscillator with a large non-linear frequency shift. In fact, the power $p$ of the injection-locked oscillator is given by $\frac{p}{p_0} = 1 + \frac{\sigma I_{ac,MAX}}{\sqrt{1+(N/\Gamma_P \xi)^2}}$, where $\xi = I_{dc}/I_{th}$ is the supercriticality of the dc bias current (or voltage) and $\Gamma_P$ is the effective damping rate [29]. For the device we have studied, $N/\Gamma_P$ is larger than 15, hence a reduced dependence of the oscillator power as a function of the $I_{dc}$ is achieved [29] as shown in Fig. 2(b) (blue diamonds) where a change of less than 3% is observed in $dm_X$.

The quasi-linear behavior of the intrinsic phase shift shown in Fig. 2(b) can be approximated by $\varphi(I_{dc}) = mI_{dc} + n$. The fitting parameters can be identified directly from the rectified voltage as follows. The $V_{dc,MAX} \cong \frac{1}{4}(R_{AP} - R_P)I_{ac,MAX}\sqrt{p}$ is obtained at $I_{dc,0}$ (see Fig. 2(a)), for this value of current the argument of the cosine is zero hence $n = -mI_{dc,0}$. In addition, the value of m can be estimated from the second derivative of the rectified voltage with respect to the dc current evaluated at $I_{dc,0}$, $\left.\frac{d^2 V_{dc}}{dI_{dc}^2}\right|_{I_{dc}=I_{dc,0}} = -m^2 V_{dc,MAX}$. This approach can be used also with experimental data to extract information about the intrinsic phase shift. Eq. 5 can be then rewritten as the equation of a parabola: $V_{dc}(I_{dc}) = aI_{dc}^2 + bI_{dc} + c$ where the expressions of the coefficients are $a = -\frac{1}{2}V_{dc,MAX}m^2$, $b = V_{dc,MAX}m^2 I_{dc,0}$, and $c = V_{dc,MAX}(1 - \frac{(mI_{dc,0})^2}{2})$. This parabolic relation, which is a key result of this work, can be used to estimate the DOR obtaining results very close to the direct parabolic fit of the data. Supplementary Fig. S1(a) [34] shows a comparison between the parabolic fit of the micromagnetic data observed in Fig. 2(a) and the parabola obtained with the analytically evaluated parameters (see Supplementary Fig. S1(b) [34]), highlighting an excellent agreement.

It is clear that, in order to implement a multiplier with spintronic diodes, we need the devices to work with currents and microwave input frequencies driving an intrinsic phase shift $\varphi$ close to 0 or $\pi$. Fig. 2(c) summarizes the results of a systematic study of $\varphi$ as a function of $I_{dc}$ and $f_{ac}$ for $I_{ac,MAX} = 70.7$ µA with the indication of the threshold current value $I_{th}$. The horizontal line is the working point for the data in Fig. 2(a) and (b). For this device geometry, the value $\varphi = 0$ is achieved close to the edge of the locking range.

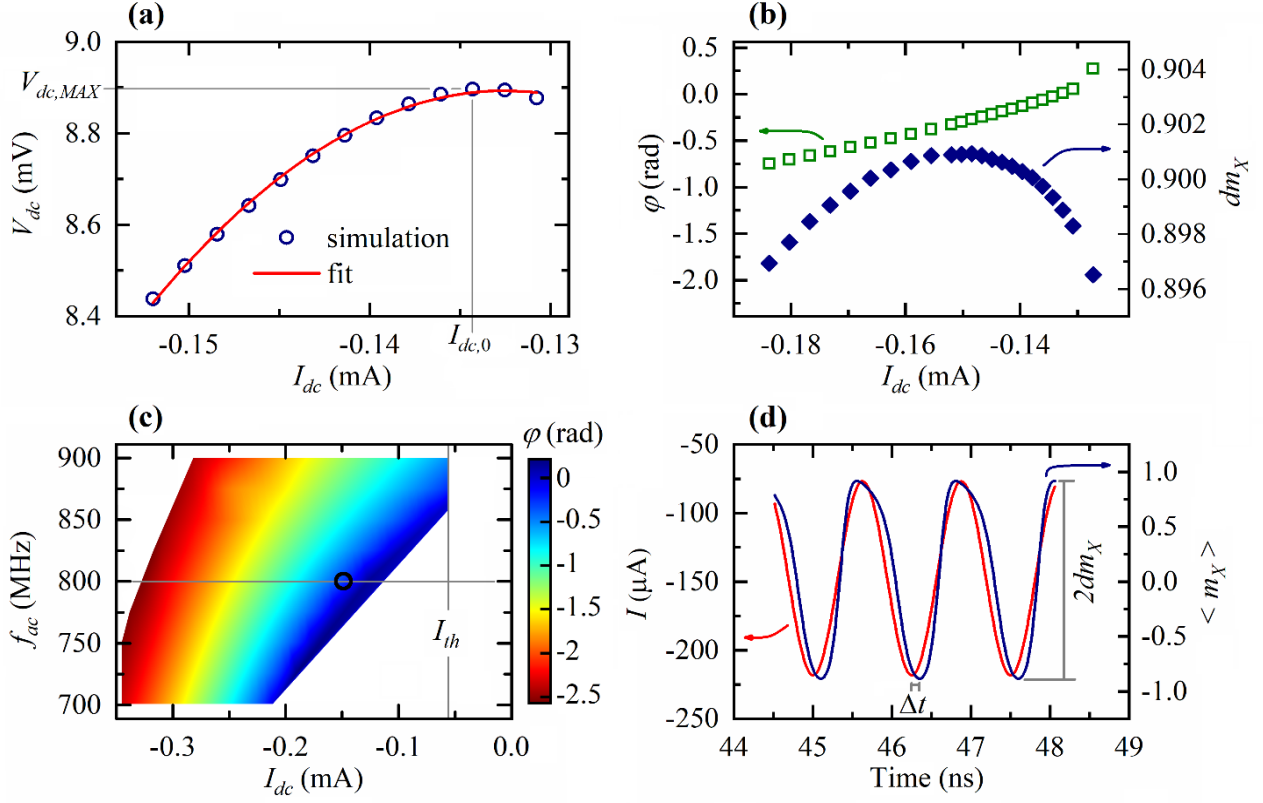

FIG. 2. (a) Plot of the rectified dc voltage as a function of the dc current applied to the spin torque diode, for $I_{ac,MAX} = 70.7$ µA and $f_{ac} = 800$ MHz, the blue circles are the results of micromagnetic simulations and the red line is the parabolic fit. (b) Intrinsic phase shift (empty squares) and amplitude of the magnetization along the x-axis (full diamonds) as a function of the dc current for the same $I_{ac,MAX}$ and $f_{ac}$ of (a). (c) Phase diagram of the intrinsic phase shift as a function of the microwave frequency and the dc current for $I_{ac,MAX} = 70.7$ µA. The vertical line is the auto-oscillation current threshold $|I_{th}| = 0.056$ mA. The horizontal line represents the microwave frequency value used for the figures in panels (a) and (b). (d) Time traces of the applied current (left y-axis) and the spatially averaged x-component of the magnetization $< m_X >$ (right y-axis) for the working point indicated with a circle in panel (c). In the figure it is also indicated the time shift $\Delta t$ between the two time traces.

Fig. 2(d) shows an example of time-domain evolution of spatially averaged x-component of the magnetization $< m_X >$ obtained for $I_{dc} = -0.148$ mA and $f_{ac} = 800$ MHz (circle in Fig. 2(c)) together with the ac current and the indication of $dm_X$.

It can be noted that a constant time shift can be identified by comparing the time traces, however, the magnetization dynamics is characterized by a first harmonic which has about 76% of the energy and high order harmonics with the other 24% (as shown in Supplementary Fig. S2 [34]) that may influence the measure of the intrinsic phase shift directly in the time domain traces. For this reason, this parameter has been computed in the Fourier space. Fig. 3 illustrates an example of the magnetization evolution when a dc current step is applied to achieve the injection-locking regime. The transient time is about 10 ns and represents a good estimation of the speed of multiplication computation.

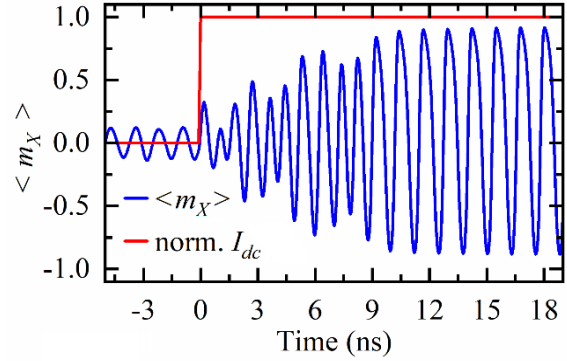

FIG. 3. An example of the time domain trace of the injection locking of the x-component of the magnetization (blue solid line) achieved after the application of a dc current step from 0 to $-0.148$ mA (normalized dc current is shown in red solid line). The ac current has an amplitude $I_{ac,MAX} = 70.7$ µA and $f_{ac} = 800$ MHz.

Supplementary Figure S3(a) [34] shows a simulation of the same device with $I_{ac,MAX} = 70.7$ µA and $f = 543$ MHz at room temperature. As can be noted, in presence of a thermal field, the frequency of the self-oscillation is smaller, as expected by the reduction of the saturation magnetization, and the time required for the locking is reduced.

Interestingly, experimental data of rectification voltage as a function of the bias current in high sensitivity spin-torque diodes also exhibits a trend similar to the theoretical curves proposed in this work for the evaluation of the DOR (see Fig. 5(e) in [15], those data are partially reported in Fig. 4 and in Supplementary Figure S3 (b) [34]). To achieve large sensitivity, there is an additional term in the rectified voltage which is proportional to the dc current and will give rise to a linear shift of the parabolic equation $V_{dc}(I_{dc}) = aI_{dc}^2 + (b + \Delta R_{dc})I_{dc} + c$, where $\Delta R_{dc}$ is the variation of the dc resistance induced by the microwave input [15]. In the rest of the paper, we will show calculations performed with the DOR estimated either from the theoretical and the experimental curves.

**IV. DOR-BASED MULTIPLICATION**

From the device input-output relationship it is possible to identify the parameters a, b, and c, which satisfy the relation $V_{dc}(I_{dc}) = aI_{dc}^2 + bI_{dc} + c$ linking the bias current $I_{dc}$ and the rectified dc voltage $V_{dc}$. The range of input currents is then scaled to the range of desired input $x$ (let's consider the range $[-1,0]$ for simplicity) with the following linear transformation $I_{dc} = |I_{dc,0} - I_{dc,-1}|x + I_{dc,0}$, where $I_{dc,0}$ and $I_{dc,-1}$ are the current values associated with the numeric input 0 and $-1$. In this way, we get an even parabolic equation and $V_{dc}(x) = V_{dc}(-x)$. The new parabolic relation is then given by $V_{dc}(x) = a'x^2 + c'$, where $a' = a|I_{dc,0} - I_{dc,-1}|^2$, and $c' = V_{dc,MAX}$ (see Fig. 4). The final calculation to multiply $F * G$ relies on the evaluation of the voltages for $x = F, G, (F - G)$ combined as described in Eq. 1.

As an example, let's consider Fig. 4, where experimental values of the rectified $V_{dc}$ are reported vs. both the $I_{dc}$ and the input $x$. If we consider $F = -0.62$, $G = -0.44$ and $F - G = -0.18$, the respective $V_{dc}$ values, $V_{dc,F} = 17.85$ mV, $V_{dc,G} = 18.30$ mV and $V_{dc,F-G} = 18.57$ mV, can be used in Eq. 1 considering the parameters $a' = -2.0476$ mV and $c' = 18.565$ mV. In this way, the product obtained is equal to 0.241 which is very close to the desired value $FG = 0.273$.

The speed of multiplication depends on which one of the two following scenarios is implemented in hardware:

1) *Max velocity*. This is achieved considering three diodes for each multiplication and the CMOS circuitry to perform the addition. The time required is the time necessary to achieve the locking plus the one necessary to perform the addition (the division is achieved together considering a proper gain for the analog adder).

2) *Minimum area occupancy*. The three DOR operations are performed with the same diode. In this case, the time increases at least three times and additional memory elements are needed to store the data before performing the sum.

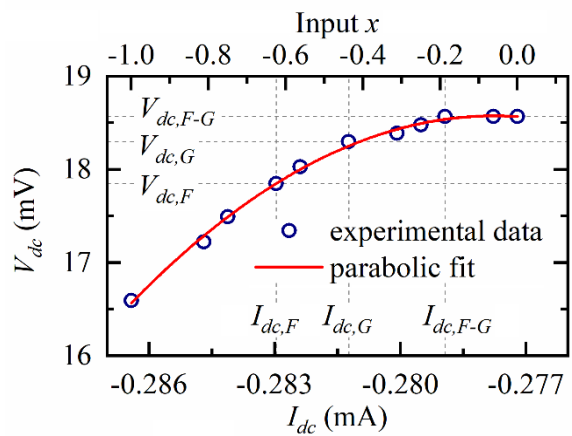

FIG. 4 – Experimental data (blue circles), from ref. [15], of the rectified voltage as a function of the bias current of the injection-locked STD (bottom axis) and the numeric input (top axis) for the even parabolic equation.

The solid red line is the parabolic fit. The points corresponding to the input values $F = -0.62$, $G = -0.44$ and $F - G = -0.18$ are also indicated.

## V. PERFORMANCE EVALUATIONS

As a first step, we have compared the micromagnetic and experimental DOR-based multiplication with the ideal one. Fig. 5(a) and (b) show 200 multiplications obtained with the DOR as computed from the numerical data in Fig. 2(a) (blue circles) and the experimental data of Fig. 4 compared to the output of the ideal multiplication (red line). The results show that the correlation between the ideal case and the micromagnetic (experimental) DOR multiplication is 99.93% (99.83%).

The second test is the calculation of the convolution between an image of a snail (extracted from the ImageNet dataset [35]) with $3 \times 3$ filters. Fig. 5(c) illustrates the probability density functions (PDFs) of the correlation coefficients $r$ computed by considering 10000 random instances of filters. The average correlation coefficients are $\bar{r}_{sim} = 99.41\%$ and $\bar{r}_{exp} = 97.87\%$ for the simulation and experimental data, respectively. The smaller average correlation and the larger dispersion achieved with the experimental data is due to a less accurate parabolic behavior. It is important to highlight that micromagnetic calculations suggest that MTJs can be very effective in the implementation of convolution reaching correlation values with the ideal multiplication larger than 99%. As an example, we show the convolution of the snail image, represented in Fig. 5(d), with a $3 \times 3$ blurring filter, consisting of equal weights. This filter is not included in the random statistical analysis of Fig. 5(c), but it represents a limit case since multiplication errors are similar and they constructively accumulate in convolution. Figs. 5(e)-(g) show the result of convolution in the case of ideal (e) and DOR ((f) simulation and (g) experimental) based multiplication. The correlation coefficients are $r_{sim} = 99.07\%$ and $r_{exp} = 96.64\%$, which, in fact, are beyond the lower tails in the PDFs of Fig. 5(c). Similar results have been also achieved with other images from the same database.

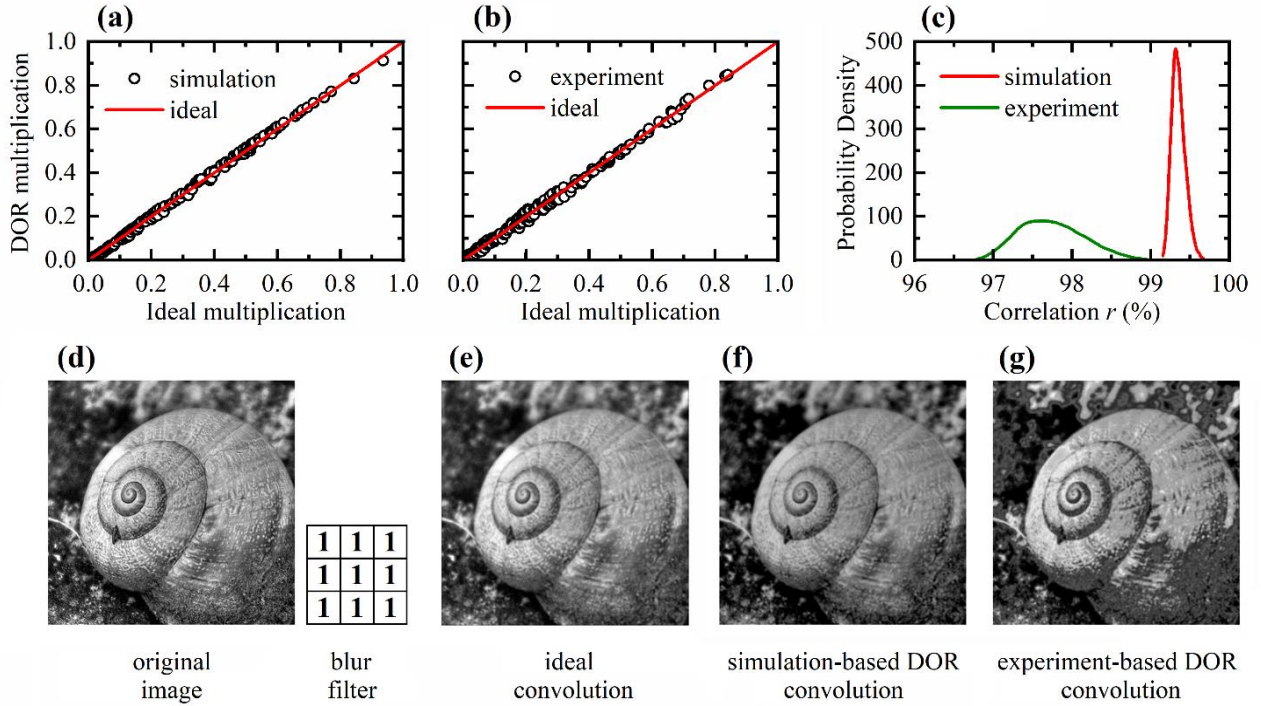

FIG. 5. (a) A comparison between DOR multiplications based on micromagnetic simulations (black circles) and the ideal multiplication (bisector of the 1$^{st}$ quadrant - red line). (b) The same comparison of (a) but with the use of the experimental curve from ref. [15] for the DOR multiplication. (c) Correlation probability density functions for the convolution of 10000 random filters considering the DOR multiplication with the simulation (red curve) and experimental data (green line). (d) Image of a snail, extracted from ImageNet dataset; in the inset, the 3 × 3 blur filter used for the convolution. (e) Ideal convolution; (f) DOR-based convolution obtained through micromagnetic data; (g) DOR-based convolution obtained through experimental data.

Today there is a growing interest in developing precision-scalable architectures [2] [36] including binary neural networks. With this in mind, we have benchmarked the impact of DOR-based multiplication in a simple CNN, using for this application the experimental data which gives the worst performance. Even if the DOR based multiplication gives rise to a reduced precision on the results, our study shows a low impact on the accuracy of neural networks.

In detail, we consider a vanilla CNN with the architecture shown in Fig. 6(a). It is composed of a single convolutional layer with 16 filters of 3 × 3 size, its output feature maps propagate through a layer of neurons having the ReLU (Rectified Linear Unit) activation function and a pooling layer that halves the results spatial dimension with the max-pooling operation. The CNN has then a flatten layer and a FC layer with 10 neurons with the softmax activation function.

The CNN is trained in Python with TensorFlow on the MNIST dataset considering a training set of 48000 images, and a validation set of 12000 images and testing is performed on a test set containing 10000 images. To prevent the overfitting dropout layers [37] and early-stopping is used. The recognition accuracy reached is 98.64% on the training set and 98.57% on the test set. Then, we use the trained weights to evaluate the accuracy on the same test set considering the DOR-based multiplication for: the convolutional layer (Conv$_{DOR}$) (the recognition accuracy achieved is 96.83%) and both the convolutional and FC layers (Conv$_{DOR}$ + FC$_{DOR}$) (94.72%), see Tab. 1 (row a). Considering that this numerical test simulates a possible effect of the hardware implementation of multiplication with STDs, the accuracy reduction ($< 4\%$) can be improved with few training iterations of the FC layer when the DOR-based multiplication is applied to the convolutional one (Conv$_{DOR}$ + trainFC); in this case we use the DOR-based convolution for the convolutional layer and the ideal multiplication for the fully connected one. After this additional training, the recognition accuracy on the test set increases to 98.40%, a value comparable to the one achieved initially (98.57%).

In addition, we simulated the device-to-device variations of the STDs. We carried out simulations adding a $\pm 2.5\%$ random variation of the parameters of the parabola performing the DOR-based multiplication (see Supplementary Note 1 [34] for more details). Tab. 1 (row b) summarizes the results obtained considering this non-ideality of the device for each of the cases described above.

Further simulations are performed considering a DOR-based multiplication computed with a parabola having a larger current region for the input (see Supplementary Figure S3(b) [34]). Tab. 1 (row c) shows the accuracies obtained with this configuration. We wish to highlight that the additional training phase is able to fix the error introduced in using this curve which is not well approximated by a parabola.

Fig. 6(c) shows some feature maps of a test image obtained with the ideal multiplication (top) and the DOR-based multiplication (bottom).

| | | Test Accuracy (%) | | |
|---|---|---|---|---|
| | Ideal | Conv$_{DOR}$ | Conv$_{DOR}$ + FC$_{DOR}$ | Conv$_{DOR}$ + trainFC |
| **a** | | 96.83 | 94.72 | 98.40 |
| **b** | 98.57 | 85.51 | 51.18 | 98.33 |
| **c** | | 97.07 | 93.11 | 98.35 |

TAB. 1: Summary of the results on the test set for CNN trained respectively with ideal multiplication (Ideal), with DOR-based multiplication applied to the convolutional layer (Conv$_{DOR}$), with DOR-based multiplication applied to both the convolutional and FC layers (Conv$_{DOR}$ + FC$_{DOR}$), and finally the results of the implementation of an additional training of the FC layer when the DOR-based multiplication is applied to the convolutional one (Conv$_{DOR}$ + trainFC). (a) Results obtained with the main curve (represented in Fig. 4 and Supplementary Figure S3(b), red curve); (b) results obtained considering device-to-device variations of the STDs; (c) results obtained using the curve with a larger input current range (Fig. S3(a), blue curve).

In the last part of this work, we have also investigated how the DOR-based multiplication can affect the estraction of the dark knowledge. Dark knowledge is at the basis of the distillation process in neural networks which has been developed primarily with the aim to transfer the knowledge from a large network model to a smaller one more suitable for deployment [23] [38]. Dark knowledge is revealed through soft probabilities, which effectively smooth out the probability distribution and reveal inter-class relationships. To obtain it, an increase of the temperature coefficient ($T$) of the softmax function $softmax(x_i) = \frac{e^{z_i/T}}{\sum_j e^{z_j/T}}$ where $z_i$ are the logits (where usually $T = 1$ is used) [23] is necessary:. Fig. 6(d) shows the output probability of the most probable class for two representative images (similar results are achieved for most of the other images) as a function of temperature while the inset shows the probability of the second most probable class as a function of temperature. It can be observed that the extraction of the dark knowledge remains robust when using the DOR-based multiplication. Fig. 6(e) summarizes the probability for all the classes for $T = 10$.

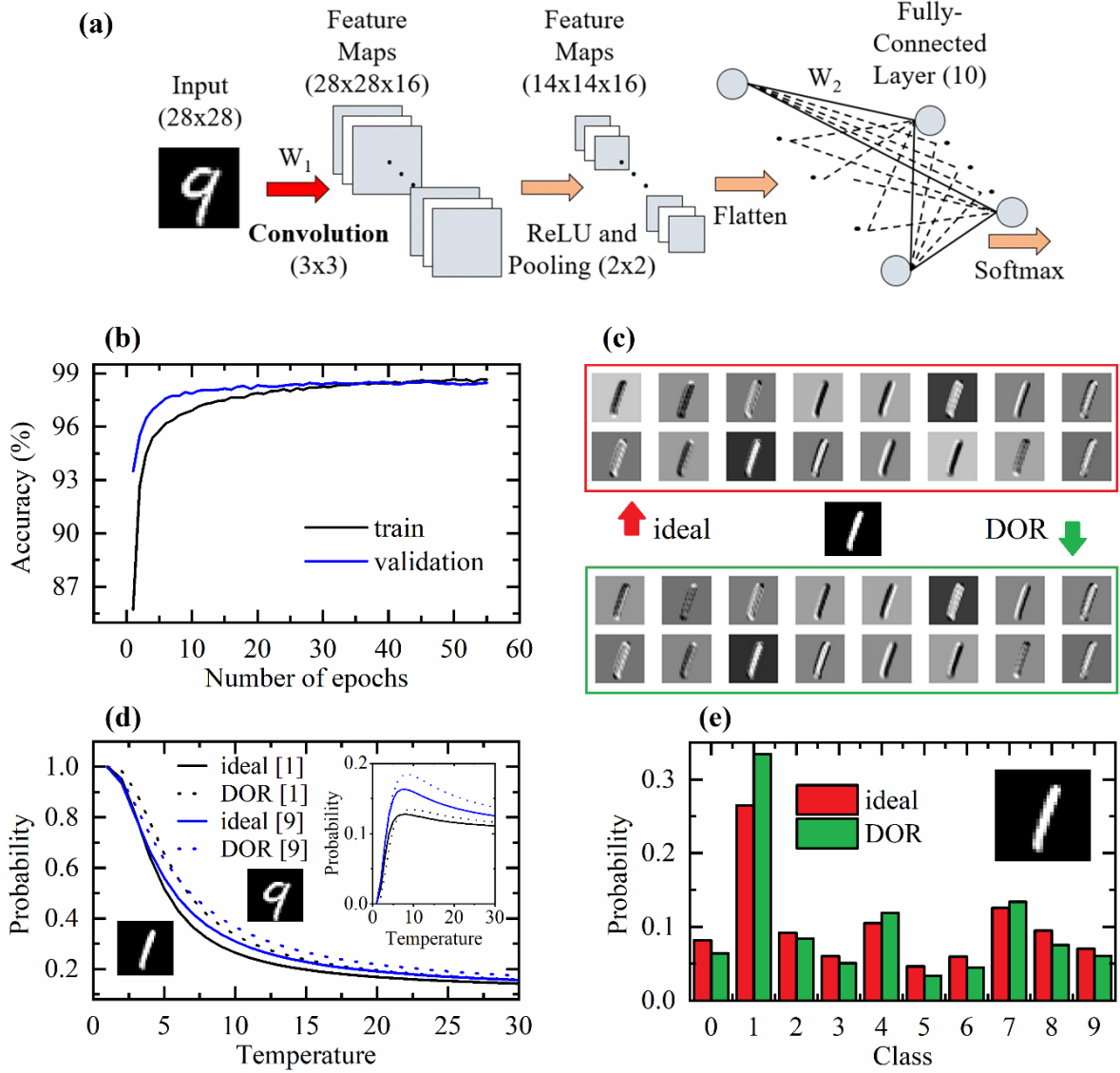

FIG. 6. (a) Structure of the CNN composed of convolutional layer, ReLU activation function, Pooling layer, FC layer of classifications. (b) Percentage of recognition accuracy as a function of the number of epochs. Black (blue) line corresponds to the results for the training (validation) dataset. (c) Feature maps of a test image obtained with ideal multiplication (top) and DOR multiplication (bottom). (d) Probability of the most probable class as a function of the temperature coefficient for the two represented test images (black line: image of handwritten digit one; blue line: image of handwritten digit nine) obtained from CNN based on ideal multiplication (solid line) and CNN with multiplication based on DOR for the convolutional layer with additional training of the FC layer (dashed line); the graph in the inset shows the probability of the second most probable class as a function of temperature. (e) Probability of all the classes for $T = 10$ for the represented image of handwritten digit one, obtained from CNN based on ideal multiplication (red) and CNN with DOR-based multiplication applied to the convolutional layer with the additional training of the FC layer (green).

## VI. SUMMARY AND CONCLUSIONS

In this work, we have shown the strategy to implement the analog multiplication with injection-locked STDs introducing the concept of DOR. In particular, those devices exhibit a parabolic rectification curve as a function of the dc current near the current region where the intrinsic phase shift is close to zero; hence the DOR, which is nothing more than the rectified voltage, can be used as an observable function for the implementation of the analog multiplication. Thanks to micromagnetic simulations and experimental data, we have shown that the DOR-based multiplication is robust, can be achieved at zero-bias field, and can be used for neuromorphic applications. In addition, future studies can take advantage of a properly designed in-plane field originated by the polarizer to have an additional degree of freedom for improving the DOR curve. Our results open a path for a low power consumption (tens of μW per operation) and potentially high-speed implementation of the multiplication operation (< 15ns) paving the way towards processing in mobile devices and IoT nodes. However, we have estimated that the performance of the approach can be improved in term of velocity and power consumption by using MTJs with lower resistance-area product [39], larger oscillation frequencies which originate faster injection locking [40] and reduced currents lowering the dissipation power per operation. Further advancements can be potentially achieved with future solutions based on antiferromagnets where the injection locking is achieved at tens of ps as predicted by micromagnetic simulations [41] [42] [43] and shown experimentally for the switching [44]. High-speed multiplication is crucial not only for neuromorphic computing but also for real-time signal processing applications.

ACKNOWLEDGEMENTS

This work was supported under Grant No. 2019-1-U.0. ("Diodi spintronici rad-hard ad elevate sensitività - DIOSPIN") funded by the Italian Space Agency (ASI) within the call "Nuove idee per la componentistica spaziale del futuro", the project PON R&I 2014-2020 No. ARS01_01215 ("New satellites generation components – NSG") funded by the Italian Ministry of University and Research and the project PRIN 2020LWPKH7 funded by the Italian Ministry of University and Research. The work has been also supported by Petaspin Association (www.petaspin.com). The authors thank Caterina Ciminelli for the fruitful discussions.

# Supplemental Material

Supplementary Figure 1

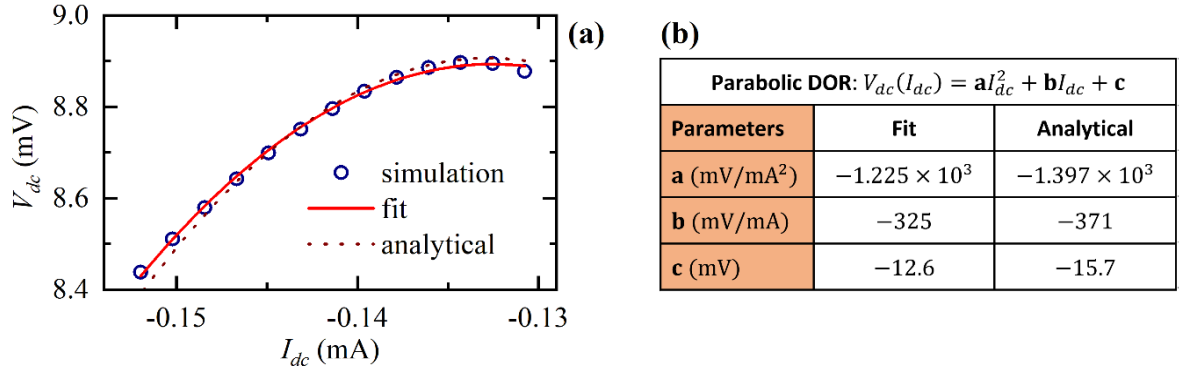

FIG. S1. Comparison between the values of $V_{dc}$ obtained from micromagnetic analyses (blue circles), the parabolic fit (red line) and the parabola obtained with the analytical data (dashed line). The table summarizes the coefficients of the two parabolas, obtained from the parabolic fit and the analytical process.

Supplementary Figure 2

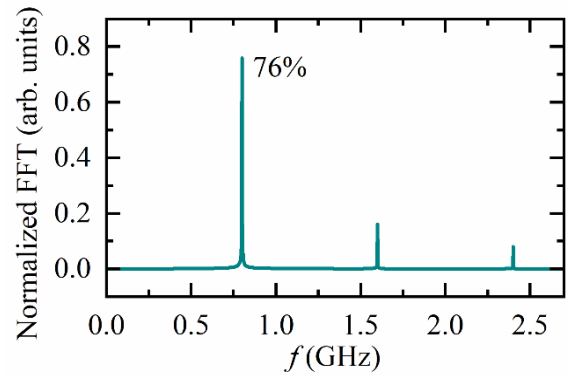

FIG. S2. FFT of the x-component of the magnetization obtained for $I_{dc} = -0.148 mA$, $I_{ac} = 70.7 \mu A$ and $f_{ac} = 800 MHz$ (represented by the circle in Fig. 2(c) of the main text), normalized by the sum of the three main peaks. The contribution of the first harmonic is about 76% of the total.

Supplementary Figure 3

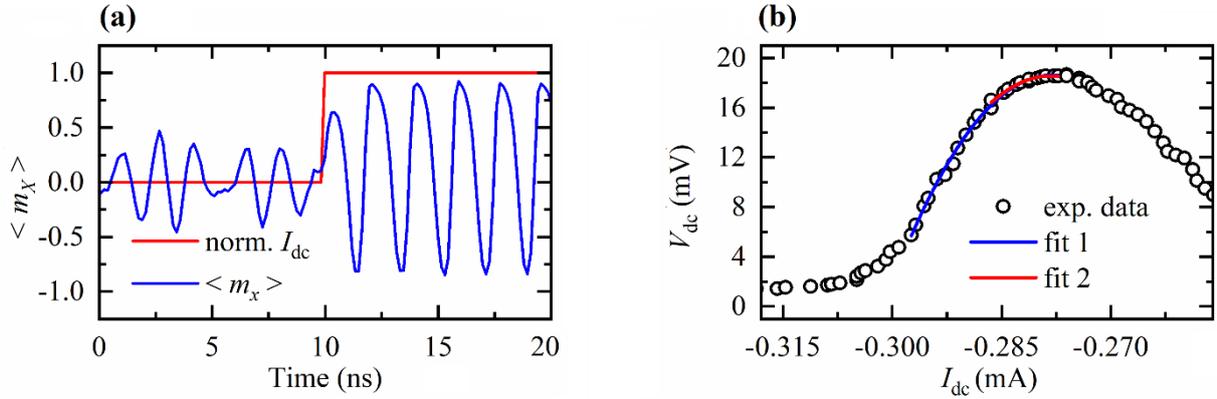

FIG. S3. (a) Example of the application of a dc current step from 0 to $-0.177$mA (normalized dc current is shown in red) and plot of the transient of the magnetization along the x-axis (blue), when the alternated current is applied with amplitude $I_{ac,MAX} = 70.7\mu A$ and $f = 543$ MHz at room temperature. (b) Experimental data obtained representing the output dc voltage depending on the input dc current in an injection locked STD (blue circles) taken from ref. [15] fitted by ideal parabolas considering a smaller (red curve) and a wider range (blue curve). Near the peak, the red curve overlaps the blue one.

Supplementary Note 1

This note describes the process used to evaluate the robustness of the system to the device-to-device variability.

We simulated the 200 random multiplications and evaluated the impact of the DOR-based multiplication in a CNN as described in the main article adding a $\pm 2.5\%$ random variation of the parameters of the parabola describing the relationship between current and voltage. The variations critically influence the accuracy of the multiplication, as it is clear to see in Fig. S4(a).

Fig. S4(b) shows the convolution between an input image and a weighted filter. The output given by the convolution operation is the sum of the products between the input and the filter; according to Eq. 1, it is possible to perform this product with the DOR, which basically uses a parabola. In Eq. 1, $F$ represents a pixel of the input image and $G$ is a weight that makes up the filter. To evaluate the robustness of the system to the device-to-device variations, we consider that, for each DOR-based multiplication between $F$ and $G$, the parabola will have a small variation which is reflected in its coefficients $a'$ and $c'$. The same reasoning was made for each multiplication of the fully connected (FC) layer when the DOR-based multiplication is applied. In summary, for each of the thousands DOR-based multiplications, we considered different parabolas adding a $\pm 2.5\%$ random variation on the parameters.

A comparison between the application with and without the variability can be observed in Tab. 1 (a,b).

We can observe a significant accuracy drop for the application of the DOR-based multiplication in the convolutional layer and in the fully connected one (Conv$_{DOR}$ + FC$_{DOR}$) in the test phase which would make the network unusable. However, few iterations of training of the FC layer (fine tuning) when the DOR-based multiplication is applied to the convolutional layer (Conv$_{DOR}$ + trainFC) lead to optimal results also with varying parameters.

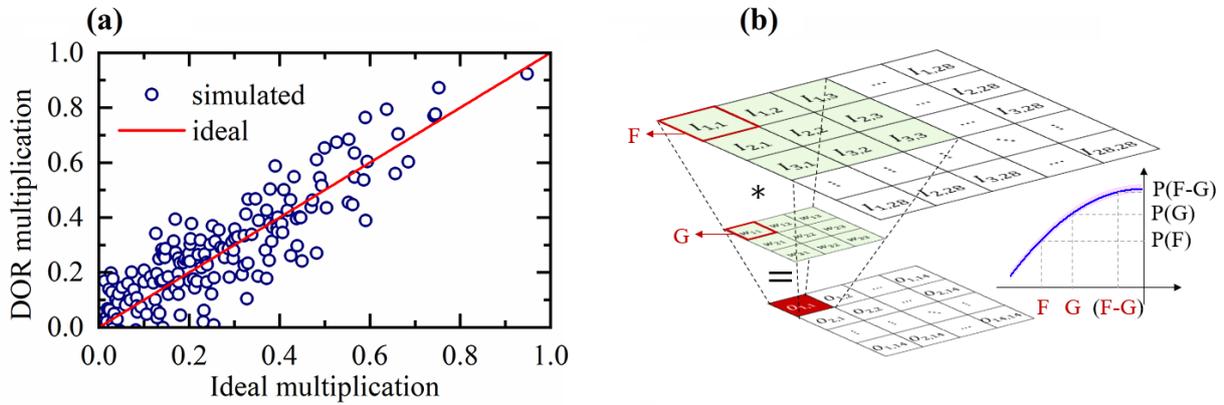

FIG. S4. (a) 200 random multiplications obtained with the DOR-based method adding a 2.5% random variation on the parameters. (b) Sketch of the convolution between an input image (top) and a weighted filter (center), where an image input pixel and filter weight are highlighted to indicate an example of how these become parabola inputs to perform the DOR-based multiplication.